\documentclass{aa}  

\usepackage[normalem]{ulem}
\usepackage{txfonts}
\usepackage{graphicx}
\usepackage{subfigure}
\usepackage{adjustbox}
\usepackage{hyperref}
\usepackage[switch]{lineno}
\hypersetup{
    colorlinks=true,
    linkcolor=blue,
    filecolor=blue,      
    urlcolor=blue,
    citecolor=blue,
    }
\usepackage[percent]{overpic}
\usepackage{xcolor}

\begin{document}

   \title{Radiation pressure role in accreting massive black hole binaries}

   \title{Radiation pressure role in accreting massive black hole binaries}


   \author{Fabiola Cocchiararo
          \inst{1,2}\fnmsep\thanks{E-mail: f.cocchiararo@campus.unimib.it}, 
          Alessia Franchini\inst{2,3}\fnmsep\thanks{E-mail: alessia.franchini@unimi.it},
          Alessandro Lupi \inst{1,2,4}, and Alberto Sesana \inst{1,2}
          }

   \institute{
    Dipartimento di Fisica "G. Occhialini", Università degli Studi di Milano-Bicocca, Piazza della Scienza 3, 20126 Milano, Italy             
    \and
    INFN, Sezione di Milano-Bicocca, Piazza della Scienza 3, 20126 Milano, Italy
    \and
    Dipartimento di Fisica "A. Pontremoli", Università degli Studi di Milano, Via Giovanni Celoria 16, 20134 Milano, Italy 
    \and
    DiSAT, Università degli Studi dell’Insubria, via Valleggio 11, I-22100 Como, Italy\\
       }

   \date{}

 
\abstract
{We investigate the impact of radiation pressure on the circumbinary discs surrounding accreting massive black hole binaries (MBHBs) at milli-parsec separations, using 3D hyper-Lagrangian resolution hydrodynamic simulations.
The circumbinary discs in our simulations evolve under an adiabatic equation of state. The gas temperature is therefore allowed to change through viscous heating, black-body cooling and self-gravity. We take a significant step further by including the contribution of radiation pressure in the simulations. We model binaries with a total mass of $10^6 \, M_{\odot}$, eccentricities $e=0,0.45,0.9$ and mass ratios $q= 0.7, 1$. We find that the radiation pressure significantly alters the vertical and thermal structure of the disc, resulting in a geometrically thinner, therefore colder configuration. This leads to a reduced accretion rate onto the binary and suppresses cavity eccentricity growth and precession in circular equal mass binaries.  
The binary eccentricity remains approximately constant, while the semi-major axis decreases over time due to net negative torque, regardless of the initial binary orbital parameters.}

   
\titlerunning{}
   \authorrunning{F. Cocchiararo et. al}
   \maketitle
%
\section{Introduction}
In recent years, the interaction between massive black hole binaries (MBHBs) and 
circumbinary discs has been extensively investigated with the aim of better understanding how the disc affects the evolution of the binary orbit.
Despite significant progress, the diversity in numerical methods (e.g. 2D vs 3D codes or Eulerian vs Lagrangian codes) and physical models adopted (e.g. locally isothermal vs adiabatic equations of state, inclusion of the disc self-gravity) has resulted in a variety of results, making it difficult to piece together a coherent picture \citep{Munoz2019,Duffell2020,tiede2020,heathnixon2020,Franchini2022}. Within this context, \cite{SantaBarbara2024} presented the results obtained using a common, simple, binary-disc setup and performing numerical simulations with a variety of hydrodynamic codes to measure, among other properties, the magnitude of the gravitational torque that the disc exerts on the binary. 
While the study shows general agreement on the sign of the gravitational torque between different codes, there are still significant differences in its absolute magnitude and in other aspects of the binary-disc evolution that essentially depend on the code geometry and on the employed boundary conditions.

The choice of a physical model for the disc greatly influences the outcome of the interaction with the binary. 
For instance, the disc aspect ratio $H/R$ was recently found to play a critical role in the binary orbital dynamics, determining whether the binary inspirals or outspirals as a result of the interaction with the gas \citep{tiede2020,heathnixon2020,Franchini2022}.
Additionally, self-gravity in circumbinary discs has been shown to regulate both the torque exerted onto the binary and disc temperature, leading the binary to shrink regardless of the disc initial temperature - i.e. initial aspect ratio - \citep{Cuadra2009,roedig2012,roedig2014,Franchini2021}. 

The vast majority of previous works used a locally isothermal equation of state \citep{farris2014,Zrake2021,DOrazio2021,Franchini2022}, assuming a constant gas temperature profile through the disc over time and therefore limiting the ability to capture shock-induced heating and its effect on the disc and ultimately on the binary.
For instance, some of the gas that leaks inside the cavity is flung back towards the inner edge of the cavity, producing shocks which can alter the disc aspect ratio and temperature profile \citep{artymowicz1996,Hayasaki2007}.

In our earlier work \citep{Cocchiararo2024}, we made a significant step further, including the gas self-gravity in the circumbinary disc together with a black-body-like cooling prescription for the gas thermodynamics evolution in 3D. We employed a live binary \citep{Franchini2023} whose orbital parameters evolve over time under the influence of gravitational and accretion torques exerted by the disc. While we focused on the characteristics of the electromagnetic emission from the disc, it is worth mentioning here that we found the interaction between the binary and the circumbinary disc to cause the binary semi-major axis to decrease with time as the disc self-regulates, reaching a stable aspect ratio value in the inner parts \citep{Franchini2021}.  We find that initially circular binaries tend towards higher eccentricity values, in agreement with previous works \citep{roedig2011,DOrazio2021,Siwek2022, Franchini2024b}, whereas very eccentric binaries experience a negligible eccentricity evolution within the time frame of our simulations.

A critical aspect neglected in previous studies is the role of radiation pressure in the hydrodynamics evolution of the circumbinary disc. In fact, at sub-parsec scales discs are likely dominated by radiation pressure; therefore, including this additional term in the hydrodynamics equations is very important in order to advance the theoretical modelling of the MBHB-disc interaction.
Radiation pressure plays an important role in the hotter inner regions of the disc, where it significantly affects the gas dynamics, potentially modifying its geometry - including its aspect ratio, cavity shape and eccentricity - and therefore altering the inflow of gas towards the binary. Correctly modelling the inflow of gas inside the cavity is fundamental as this affects the gravitational and accretion torques on the binary, the accretion rate and therefore the evolution of the binary orbital parameters, i.e. semi-major axis and eccentricity. 

In this work, we evaluate the impact of radiation pressure on the evolution of circumbinary discs around MBHBs for different values of the binary initial mass ratio and eccentricity. Building on our previous simulations \citep{Cocchiararo2024}, we implement the contribution of the radiation pressure in our 3D numerical simulations with hyper-Lagrangian refinement and consider milli-parsec scale binaries. The simulations also account for the thermodynamics evolution of the gas using a radiative cooling prescription in the form of black-body radiation. We have also included the self-gravity of the disc and the Shakura-Sunyaev prescription for viscosity \citep{ShakuraSunyaev1973}. 

This work is the first part of a larger project focusing on the role of radiation pressure in the evolution and emission of massive black hole binaries at milli-parsec separation. We hereinvestigate its impact on the binary and circumbinary disc dynamics. 
We will dedicate a follow-up paper \cite{Cocchiararo-stub-EM} to the radiation pressure influence on the electromagnetic emission and the observational signatures of these systems.

We explored the time evolution of the binary and disc properties for three values of the binary eccentricity $e=0, 0.45, 0.9$ and mass ratio $q=0.7,1$. We compare the evolution of the disc and binary orbital parameters with the simulations that included only gas pressure \citep{Cocchiararo2024} in order to isolate the effect of radiation pressure.

The paper is organised as follows. In Sect. \ref{sec:NumPhyModel}, we describe the numerical details of the simulations, how we implemented the radiation contribution to the total pressure, the physical parameters we use, and the circumbinary disc assumptions. We show and discuss the main results we obtain, including the time evolution of the main binary and disc properties in Sect. \ref{sec:results}. Finally, we report our conclusions in Sect. \ref{sec:conclusions}.


\section{Numerical and physical model} 
\label{sec:NumPhyModel}

We use the 3D meshless finite mass (MFM) version of the code {\sc gizmo}
\citep{Hopkins2015} to model the interaction between the MBHB and the circumbinary disc. We employ an adaptive hyper-Lagrangian refinement, i.e. particle splitting, to achieve a higher resolution inside the cavity carved by the binary \citep[see][for details]{Franchini2022}. We use the same refinement scheme used in our previous paper \citep{Cocchiararo2024}.
In code units, the total mass of the binary is $M_{\rm B} = M_{\rm 1} + M_{\rm 2} = 1$ where $M_{1}$ and $M_{2}$ are the masses of the primary and secondary black hole respectively, and the initial semi-major axis is $a_{\rm 0}=1$. 
The two binary components are modelled as sink particles with accretion radius $R_{\rm sink} = 0.05a_0$. As the binary orbit evolves with time \citep{Franchini2023}, mass, linear, and angular momentum are conserved during each accretion event, following the method used in the {\sc phantom} code \citep{bate1995}.
We sample the gas in the circumbinary disc with $N=10^6$ particles for a total disc mass $M_{\rm D} = 0.01 M_{B}$. This mass is distributed with an initial surface density profile $\Sigma \propto R^{-1}$ between $R_{\rm in}=2a$ and $R_{\rm out}=10a$ in the circular cases, and between $R_{\rm in}=3a$ and $R_{\rm out}=10a$ in the eccentric cases. The disc is coplanar with the binary orbit and has an initial aspect ratio $H/R = 0.1$. We generate the initial conditions of the disc using the SPH code {\sc phantom} \citep{Price2017}.

The thermodynamics evolution of the gas is governed by an adiabatic equation of state with index $\gamma = 5/3$. The gas is therefore allowed to heat and cool so that we can capture the effect of shocks. 
We model the angular momentum transport within the disc using the $\alpha$ viscosity parameter, which constitutes a simple parameterisation of the turbulence within the disc \citep{ShakuraSunyaev1973}. 
We included viscosity through the Shakura-Sunyaev parametrisation with kinematic viscosity $\nu = \alpha c_{\rm s } H$, where $c_{\rm s}$ is the gas sound speed, $\alpha=0.1$, and the disc thickness H is $H=c_{s,i}/\Omega_{K}$. 
We also include the gas self-gravity \citep{lodato2007sg} and a black-body-like cooling.
We start our discs in gravitational equilibrium by setting the initial Toomre parameter $Q > 1$ \citep{Toomre1964}.

We assume radiative cooling in the form of black-body radiation. The cooling rate is given by: 
\begin{equation}
        \label{eqn:cooling}
        \Lambda_{\rm cool} = \frac{8}{3} \frac{\sigma_{\rm SB} T^4}{\kappa \Sigma}
\end{equation}
where $\sigma_{\rm SB}$ is the Stephan-Boltzmann constant, $\kappa$ is the opacity, $\Sigma$ is the disc surface density and $ T$ is the temperature.
We assume that the opacity $\kappa$ is a combination of the Kramer's opacity $ \kappa_{\rm Kramer} \propto \rho T^{-7/2} $ and the electron scattering opacity $ \kappa_{\rm es} = 0.2 (1+X) \, \mbox{g}\, \mbox{cm}^2$, with a hydrogen mass fraction $ X=0.59$.
Since we introduce a black-body cooling, we need to choose meaningful physical units for our simulations. The total binary mass is $M_{\rm B}= 10^6 M_{\odot}$ and the initial separation is $a_0=4.8 \times 10^{-4} \, \rm{pc} \simeq 1.2 \times 10^4 \, R_{\rm g}$, where $R_{\rm g} = GM_{\rm B}/c^2$ represents the gravitational radius of the binary. 

We implement the radiation pressure contribution in {\sc gizmo} in the following way.
We first determined the initial temperature $T_{\rm 0, i}$ by solving the following relation : 
\begin{equation}
    \label{eqT}
    \frac{4}{3} \frac{\sigma_{SB}}{c} T_{\rm 0, i}^4 + \frac{\rho_{\rm i} k_{B}}{\mu m_{p}} T_{\rm 0, i}-\rho_{\rm i} c_{\rm s, i}^2 = 0 
\end{equation}
where $c_{\rm s, i} $ is the sound speed of the gas element $i$ determined by the radial profile

\begin{equation}
    c_{\rm s, i} = \frac{H}{R}\sqrt{\frac{M_{\rm 1}+ M_{\rm 2}}{a_0}}r^{-0.5}
\end{equation}
where $H/R$ is the disc aspect ratio, $M_{1}$ and $M_{2}$ are the masses of the primary and secondary black holes, respectively, $a_{0}$ is the initial semi-major axis, and $r$ is the radius in units of $a_{0}$. 

Using $T_{\rm 0, i}$, we calculated the initial internal energy $u_{\rm 0, i}$ for each gas element $i$ as the sum of the gas internal energy and the radiation internal energy. The values of $u_{\rm 0, i}$ are included in the initial conditions file, ensuring that the contribution of the radiation pressure is accounted for at the beginning of each simulation. 
We then evolve the internal energy equation over time.

At each time-step of the simulation, the temperature $T_{\rm i}$ is re-calculated from the updated internal energy $u_{\rm i}$ by solving:

\begin{equation}
    4\frac{\sigma_{\rm SB}}{c\rho_{\rm i}}T_{\rm i}^4 + \frac{3}{2}\frac{k_{\rm B}}{\mu m_{\rm p}}T_{\rm i} - u_{\rm i} = 0
\end{equation}
and the ideal gas pressure is computed as $P_{\rm gas} = (\rho_{\rm i}k_{\rm B}T_{\rm i})/(\mu m_{\rm p})$. 
In order to account for the contribution of radiation pressure, we compute the parameter $\beta$, defined as the ratio of gas pressure to total pressure: 

\begin{equation}
    \beta =\frac{1}{1+\frac{4}{3}\frac{\sigma_{\rm SB}}{c}\frac{T^4}{P_{\rm gas}}} .
\end{equation}
We use this parameter to modify and update the effective adiabatic index $\gamma_{\rm \beta}$ in the equation of state, ensuring a self-consistent treatment of the coupled gas and radiation pressure, using the following definition: 

\begin{equation}
    \label{eqn:adiabaticindex}
    \gamma_{\beta} = \beta + \frac{(4-3\beta)^2+(\gamma-1)}{\beta+12(\gamma-1)(1-\beta)}.
\end{equation} 

We run six different simulations, three of which including the implementation of radiation pressure, with the following combinations of binary eccentricity and mass ratio: 
$e=0$ and $q=1$, $e=0.45$ and $q=0.7$, $e=0.9$ and $q=1$. We run the equal mass binary simulations for 1300 binary orbits and the unequal mass simulation for 1600 binary orbits to ensure that the disc reaches a quasi-steady state, i.e. the ratio between the change in binary angular momentum over accretion rate does not change with time.

\begin{figure}[h]
    \begin{center}
    \begin{overpic}[width=0.9\columnwidth, trim=0cm 5.5cm 1.3cm 0cm, clip]{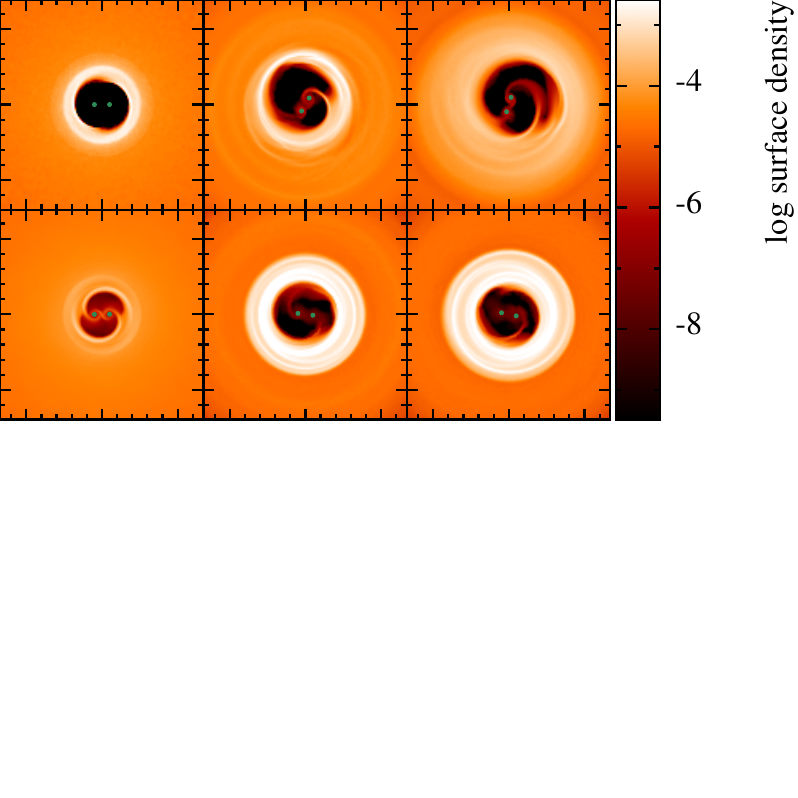}
    \put(2,60){\color{yellow}\Large e=0, q=1}
    \put(5,66){\color{black}\Large $\sim$ 1 Pb}
    \put(32,66){\color{black}\Large $\sim$ 900 Pb}
    \put(60,66){\color{black}\Large $\sim$ 1300 Pb}
    \put(-8,45){\rotatebox{90}{\color{black} \Large $P_{\rm gas}$ }}
    \put(-8,7){\rotatebox{90}{\color{black} \Large $P_{\rm gas} + P_{\rm rad}$ }}
    \end{overpic} \\
    
    \begin{overpic}[width=0.9\columnwidth, trim=0cm 5.5cm 1.3cm 0cm, clip]{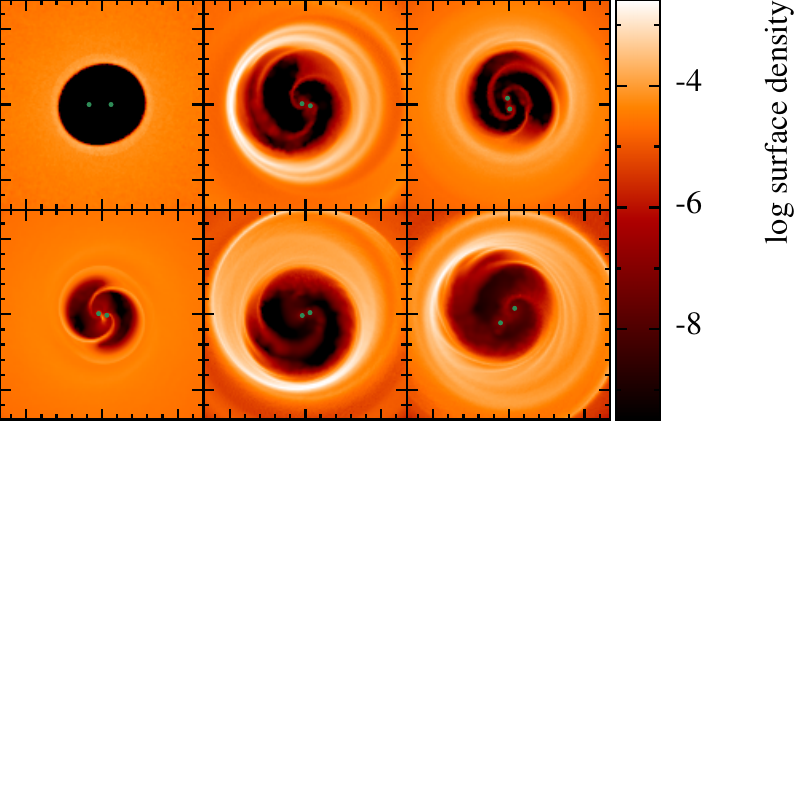}
    \put(2,60){\color{yellow}\Large e=0.45, q=0.7}
    \put(60,66){\color{black}\Large $\sim$ 1600 Pb}
    \put(-8,45){\rotatebox{90}{\color{black} \Large $P_{\rm gas}$ }}
    \put(-8,7){\rotatebox{90}{\color{black} \Large $P_{\rm gas} + P_{\rm rad}$ }}
    \end{overpic} \\
    
    \begin{overpic}[width=0.9\columnwidth, trim=0cm 6cm 1.3cm 0cm, clip]{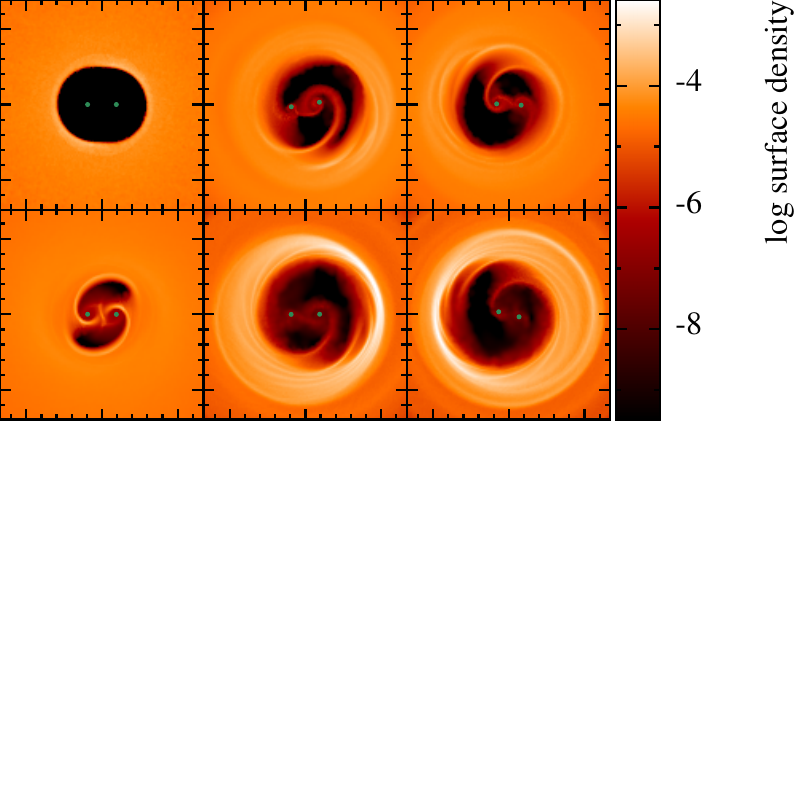}
    \put(1.2,56){\color{yellow}\Large e=0.9, q=1}
    \put(60,62){\color{black}\Large $\sim$ 1300 Pb}
    \put(-8,45){\rotatebox{90}{\color{black} \Large $P_{\rm gas}$ }}
    \put(-8,6){\rotatebox{90}{\color{black} \Large $P_{\rm gas} + P_{\rm rad}$ }}
    \end{overpic} \\

    \caption{Surface density map in code units in the $x-y$ plane for the simulations with e=0 q=1 (upper panel), e=0.45 q=0.7 (middle panel), and e=0.9 q=1 (bottom panel), at three different moments: $t=1, \,900, \, 1300 \, P_{\rm B}$ for the equal mass cases and $t=1, \,900, \, 1600 \, P_{\rm B}$  for the unequal mass case. In each panel, the first and the second row show results for $P_{\rm tot}=P_{\rm gas}$ and $P_{\rm tot}=P_{\rm gas}+P_{\rm rad}$, respectively. In both the eccentric simulations, the initial cavity radius is $3a_{\rm 0}$, while in the circular binary case it is $2a_{\rm 0}$. When the radiation pressure is included, the binary begins accreting at the very early stages of its evolution, otherwise the disc experiences a transient phase lasting approximately $900 \, P_{\rm B}$, $700 \, P_{\rm B}$ and $600 \, P_{\rm B}$ for the case e=0 q=1, e=0.45 q=0.7 and e=0.9 q=1, respectively.
    }
    \label{fig:disc}
    \end{center}
\end{figure}


\begin{figure*}
    \begin{center}
    \includegraphics[width=1.9\columnwidth ]{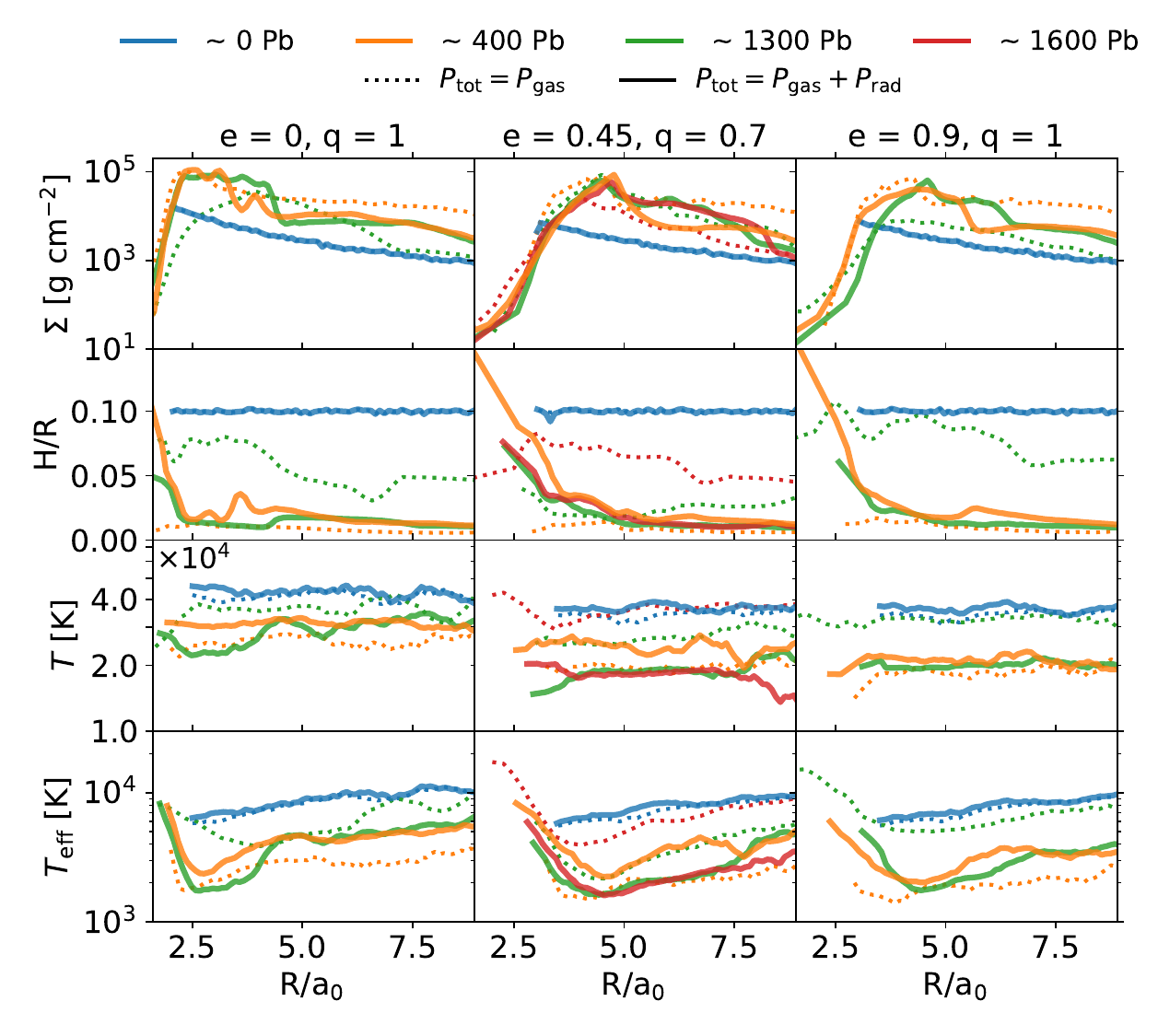} \\
    \caption{Time evolution of the surface density (first row), disc aspect ratio H/R (second row), the midplane temperature (third row), and the effective temperature (last row) as a function of radius for the binary case $e=0$ $q=1$ (first column), $e=0.45$ $q=0.7$ (central column), and $e=0.9$ $q=1$ (last column), at different times distinguished by different colours. The solid and dashed lines refers to the simulation without and with the implementation of the radiation pressure, respectively. The initial condition of the disc is the the same across all the cases, except the initial cavity radius that is $3a_{\rm 0}$ in both the eccentric simulations, while it is $2a_{\rm 0}$ in the circular binary cases. 
    We calculate the midplane temperature and the effective temperature assuming that both the gas and the radiation pressure contribute to the hydrostatic equilibrium of the disc in all the cases. The radiation pressure contribution leads the disc to maintain a lower aspect ratio and  temperature.}
    \label{fig:hr_teff}  
    \end{center}
\end{figure*}
\begin{figure}
    \begin{center}
    \includegraphics[width=0.82\columnwidth, trim=0.5cm 0.5cm 0.6cm 2cm, clip]{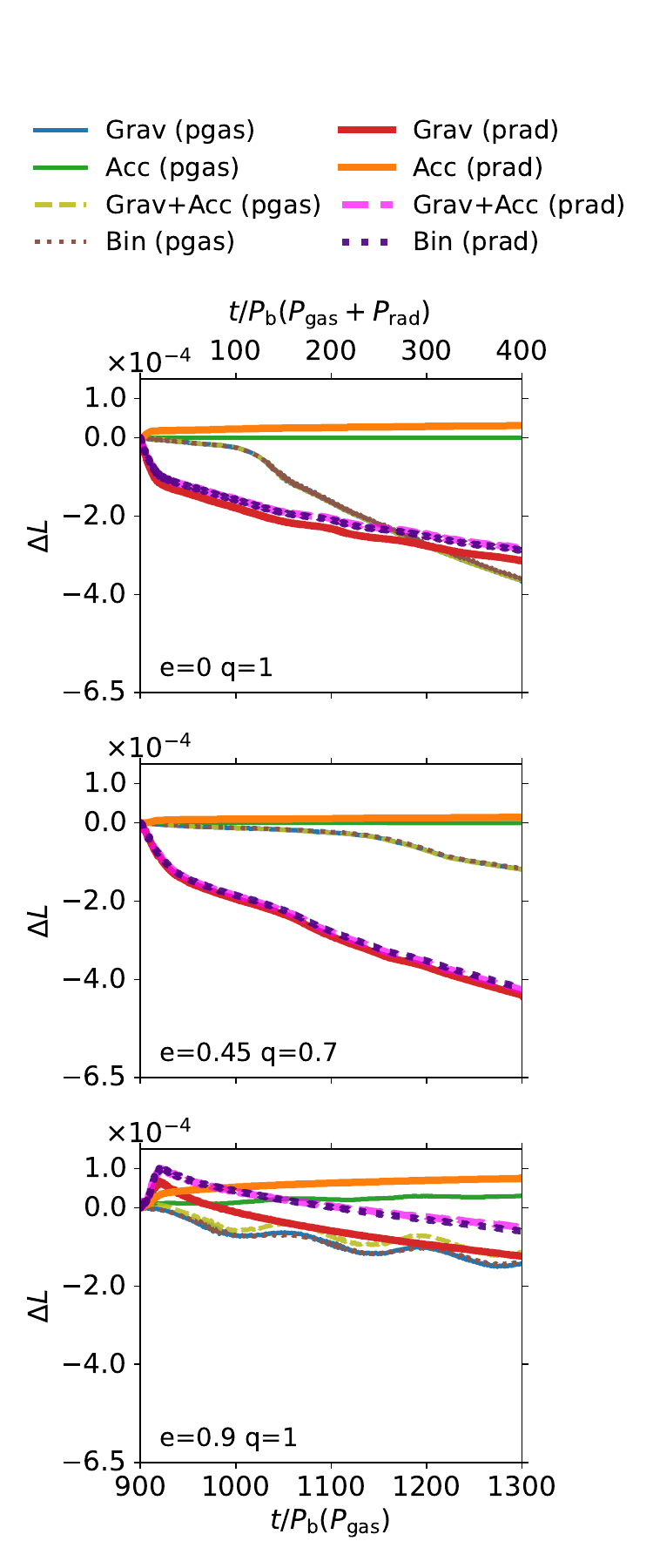}
    \caption{Angular momentum conservation for the $e=0$ $q=1$ binaries (top panel), $e=0.45$ $q=0.7$ binaries (centre panel), and $e=0.9$ $q=1$ binaries (bottom panel) over the first and last $400 \, P_{\rm B}$ for the radiation pressure and non-radiation simulations, respectively, corresponding to the time when the disc approaches a stable configuration. For simulations without the radiation pressure, the blue and dark green lines show the gravity and accretion contributions, respectively. Their sum is given by the dashed light green line. The dotted brown line represents the binary angular momentum as calculated from the simulation. For runs including radiation pressure the same quantities are plotter in red, orange,pink and purple.}
    \label{fig:momAng} 
    \end{center}
\end{figure}

\section{Results}
\label{sec:results}

We here present the results that we obtained from our numerical simulations, neglecting and including the radiation pressure contribution as explained in Section \ref{sec:NumPhyModel}. In particular, we compare the evolution of the disc properties, such as the time evolution of the disc surface density, temperature, and aspect ratio. We then compare the evolution of the binary orbital parameters and accretion rate. 
Finally, we focus on the interaction between disc morphology, the alternation of the accretion rate onto each black hole, and the preferential accretion in binary systems.

\subsection{Disc evolution}
\label{DiscEvolution}
For a more comprehensive physical understanding of circumbinary discs evolution, it is essential to account for radiation pressure, which, while is often included in GRMHD simulations \citep{sadowski2016, mishra2016, Cattorini2024, Tiwari2025}, is usually neglected in studies of binaries at large separations. 
Its effect is particularly important in the hot inner regions of the disc, where it can significantly alter the gas dynamics, the disc geometry, and ultimately influence the inflow of material toward the binary. 
Figure \ref{fig:disc} shows the surface density maps in the $x-y$ plane for the simulations with $e=0$ and $q=1$ (upper panel), $e=0.45$ and  $q=0.7$ (middle panel), and $e=0.9$ and $q=1$ (bottom panel) respectively, at $t=1, \,900, \, 1300 \, P_{\rm B}$ for the un-equal mass binary, and $t=1, \,900, \, 1600 \, P_{\rm B}$ for the equal mass binaries. In order to highlight the radiation pressure effect, we show, in each of the three panels, the same simulation which includes only the gas pressure contribution in the upper row.

The first noticeable effect of radiation pressure is to inhibit the cavity eccentricity growth, particularly in the circular equal mass binary case. 
The simulation with the circular equal mass binary and the implementation of radiation pressure shows a region of enhanced density in the inner part of the disc that extends up to $R\sim 5a$. This feature persists also after $t=1300 \, P_{\rm B}$. 
In the eccentric binary cases, the presence of radiation pressure results in a denser over-density at the cavity inner edge. This feature, called "lump", has been previously found in the literature in circumbinary discs around close-to-equal-mass binaries and it is the result of the combination of the flung-back streams impacting on the cavity edge and of the disc eccentricity \citep{Shi2012,farris2014,Noble2021,Franchini2024a}.
In the radiation pressure case, the dense region that forms in the equal mass circular binary case possibly hinders the formation of the lump. 

With our 3D simulations that include time dependent thermodynamics, we are able to resolve shocks arising from gas streams that are flung back into the cavity inner edge. 
These shocks do indeed affect the inner disc thickness and temperature. The radiation pressure contribution can also play a role in this process, potentially influencing the disc structure.

Figure \ref{fig:hr_teff} 
shows the time evolution of the surface density, disc aspect ratio $H/R$,\footnote{For each radial bin with $N$ particles, we compute the aspect ratio as $\frac{H}{R}= \frac{1}{R}\sqrt{\frac{1}{N-1}\sum_{i=1}^{N}(z_{\rm i}-\bar{z})^2}$, with $\bar{z} = \frac{1}{N} \sum_{i=1}^N z_i$ the mean of the vertical coordinates and $R = \frac{1}{N} \sum_{i=1}^N R_i$ the mean radius of the particles in each radial bin. In order to exclude gas particles too far from the disc, we select only gas particles whose density $\rho_{\rm i}$ is higher than the minimum particle density at the initial time. }
the midplane temperature, and the effective temperature $T_{\rm eff}$ as a function of the radius $R$ for all our simulations, at four different times. 
In all simulations, for each gas particle $i$, we obtained the temperature $T_{i}$ numerically solving the following implicit equation for $T_{i}$ 

\begin{equation}
    \label{eqn:Ptot}
    P_{\rm tot} = P_{\rm gas} + P_{\rm rad} = \frac{\rho k_{\rm B} T_{\rm i}}{m_{\rm p}\mu} + \frac{4}{3}\frac{\sigma_{\rm SB} T_{\rm i}^4}{c}.
\end{equation}

We divide the disc temperature domain into a 2D matrix in the x-y plane, corresponding to the binary orbital plane. For each pixel, the midplane temperature is calculated by averaging the temperatures of all the particles within the vertical range $-0.05a < z < 0.05a$, obtaining the midplane temperature matrix $T$. Finally, we calculate the effective temperature in the optically thick approximation ($\kappa\Sigma > 1 $) as: 

\begin{equation}
    \label{eqn:Teff}
    T_{\rm eff}^4 = \frac{4}{3}\frac{T^4}{\kappa \Sigma}
\end{equation}
where $\Sigma$ is the surface density of each element of the matrix.

Since the Toomre parameter is $Q>1$, the disc tends to cool down in order to reach a self-regulated state with $Q\sim1$.
Regardless of the inclusion of the radiation pressure contribution, during the first $\sim 400 \, P_{\rm B}$, the cooling leads to a decrease in the disc aspect ratio  to $H/R \sim M_{\rm B}/M_{\rm D} \sim 0.01$.
When radiation pressure is neglected, the disc still experiences this cooling phase but the aspect ratio increases again after 900 orbits, settling at a much higher value, around 0.07. This evolution of the disc aspect ratio is driven by the change in midplane temperature of the disc. As we can see from the third raw of Figure \ref{fig:hr_teff}, the midplane temperature initially decreases and then increases again, causing the vertical expansion of the disc.
On the contrary, if radiation pressure is included, the midplane temperature decreases and does not increase again.  The disc aspect ratio therefore remains at around $H/R\sim 0.02$, except for the very inner parts of the disc that are influenced by the effect of the stream shocks. Since the disc is initially completely radiation pressure dominated, the cooling dominates over heating, causing the disc vertical collapse. This behaviour is consistent with what was found in \cite{sadowski2016} and in \cite{mishra2016}. 
During this cooling phase, the surface density increases in both the radiation pressure and gas pressure only runs and regardless of the binary mass ratio and eccentricity. 
In the simulations that include the radiation pressure effect, the surface density reaches a stable profile on a shorter timescale.
This possibly implies that the additional vertical support provided by the radiation pressure rapidly regulates both the thermal balance and geometrical structure of the disc.
The effective temperature decreases from initial values between $T_{\rm eff} \sim 6 \times 10^3-10^4 \, K$ to $T_{\rm eff} \sim 1.5-7 \times 10^3 \, K$, the hottest region lying, as expected, within $R \lesssim 2.5-3 \,a_{0}$. The final effective temperature of the disc inner edge in the simulations with radiation pressure is $\sim 2-3 \times 10^3 K$, a factor at most 3 lower than the simulations that neglect the radiation pressure contribution.
Consistently with the evolution of the disc aspect ratio, the effective temperature remains lower in the simulations that include the effect of radiation pressure. 

\subsection{Binary evolution}
\label{BinaryEvolution}  

In the previous section we have shown that the radiation pressure has a non-negligible effect on the thermal properties of the disc inner edge, which ultimately regulates how much material is able to enter the cavity and accrete onto the binary.
The radiation pressure contribution to the hydrostatic equilibrium of the disc does also affect the migration of material from the outer to the inner regions of the circumbinary disc. This could result in a different contribution of the accretion torque $\mathbf{L_{\rm acc}}$ to the total angular momentum equation, potentially changing the binary fate. Indeed, conservation of angular momentum in our simulations implies that 

\begin{equation}
    \frac{d \mathbf{L}}{dt}=\mathbf{T}_{\rm G} + \frac{d \mathbf{L_{\rm acc}}}{dt},
\end{equation}
where $\mathbf{L}$ is the total binary angular momentum. The contribution from the accretion of gas particles onto the two binary components - $\mathbf{L_{\rm acc}}$ - and the contribution of the gravitational torque exerted by the disc elements onto each individual MBH - $\mathbf{T}_{\rm G}$ -, determine whether the binary orbit shrinks or expands as a result of the interaction with the circumbinary disc \citep{roedig2012}. 

We compute the gravitational torque exerted by N gas particles on each binary component as 

\begin{equation}
    \mathbf{T}_{\rm G} = \sum_{k=1}^{2} \sum_{i=1}^{N} \mathbf{r}_{\rm k} \times \frac{GM_{\rm k}m_{\rm i}(\mathbf{r_{\rm i}}-\mathbf{r}_{\rm k})}{|\mathbf{r_{\rm i}}-\mathbf{r}_{\rm k}|^3},
    \label{eqn:Tg}
\end{equation}
where $m_i$ and $\mathbf{r_i}$ are the the mass and the position of the gas particle $i$, $M_{k}$ and $\mathbf{r_k}$ are the mass and the position of the sink particle $k$. Both the gas particles and MBHs position vectors are computed with respect to the centre of mass of the system. 
The accretion torque can be computed as  

\begin{equation}
    d\mathbf{L}_{\rm acc} = \mathbf{r}_{i} \times m_{\rm i}\mathbf{v}_{\rm i} - \frac{m_{\rm i}M_{\rm k}}{m_{\rm i}+M_{\rm k}}[(\mathbf{r}_{\rm i}-\mathbf{r}_{\rm k})\times (\mathbf{v}_{\rm i}-\mathbf{v}_{\rm k})],
\end{equation}
where $\mathbf{r_{i}}$, $\mathbf{v_{i}}$, and $m_{i}$ are the position, velocity and mass of the accreted particle $i$, while $\mathbf{r_{k}}$, $\mathbf{v_{k}}$, and $M_{k}$ are the position, velocity and mass of the sink $k$. 
We calculate the cumulative variation of the binary angular momentum over the entire duration of the simulations as
\begin{equation}
    \Delta \mathbf{L} = \sum_{dt} (\mathbf{T}_{\rm G}dt+d\mathbf{L}_{\rm acc}),
\end{equation}
where $dt$ is the time step between two consecutive snapshots. 
Figure \ref{fig:momAng} shows the cumulative change in angular momentum in our six simulations. 
We note that we can self-consistently prove that we conserve the angular momentum as we evolve a live binary and are therefore able to track its orbital parameters evolution.
Regardless of the binary properties or the implementation of radiation pressure, the total angular momentum is conserved, with only minor fluctuations in the $e=0.9$ $q=1$ binary case. 
In all the simulations, the negative gravitational torque dominates and the total angular momentum variation is $\Delta \mathbf{L} < 0$, which means that the binary undergoes orbital contraction. 
In particular, in the circular case, when only gas pressure is considered, the accretion torque contribution to the total angular momentum evolution is nearly negligible. 
However, in the radiation pressure run, $\Delta \mathbf{L_{\rm acc}} > 0$, but this positive contribution is not sufficient to counterbalance the stronger negative gravitational torque, resulting in net angular momentum loss. 
We find a similar evolution in the medium eccentric case $e=0.45$. In the more eccentric case $e=0.9$, the behaviour becomes more complex. In the gas-pressure-only simulation, we observe a similar evolution to the less eccentric cases. The inclusion of radiation pressure results in a strongly positive accretion torque. As a result, the total torque becomes positive over the first $200$ orbital periods. The gravitational torque contribution experiences an initial transient phase, starting as a positive torque and then transiting to negative values that contribute to the binary inspiral.

\begin{figure}
    \begin{center}
    \includegraphics[width=\columnwidth]{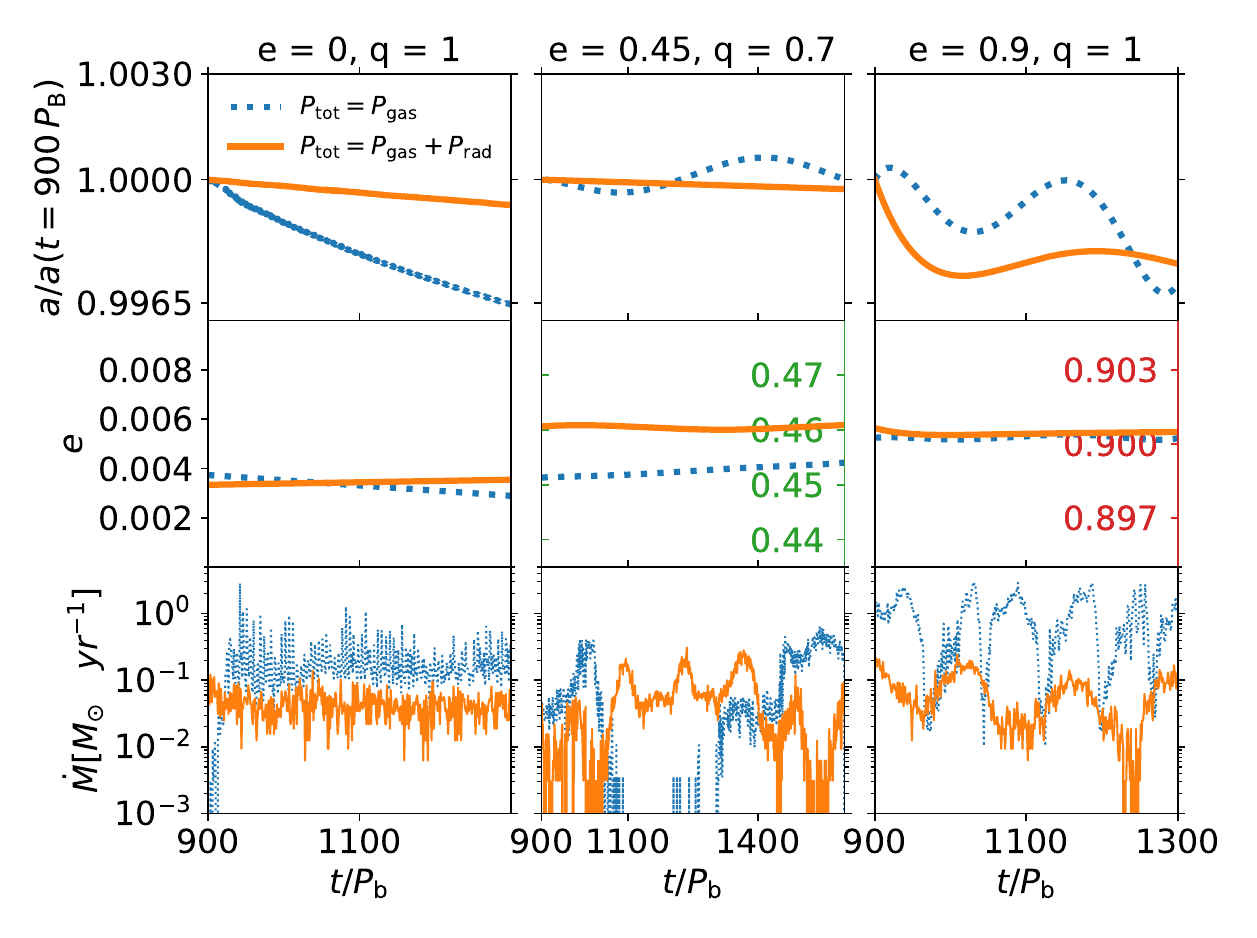} \\
    \caption{
    Time evolution of the semi-major axis (top panel), eccentricity (centre panel) and accretion rate (bottom panel) over the last 400 orbital periods for the binary case e=0 q=1 (first column) and e=0.9 q=1 (last column), and over the last 700 orbital periods for the binary case e=0.45 q=0.7 (central column). The orange and the blue lines refer to the simulation with and without the radiation pressure contribution, respectively. The semi-major axis is normalised to set $a(t=900 \, P_{\rm B})=1$. since this is the point where our binaries start to accrete material. 
    }
    
    \label{fig:a_e_mdot}
    \end{center}
\end{figure}

Figure \ref{fig:a_e_mdot} shows the time evolution of the semi-major axis, eccentricity and accretion rate over time for all our simulations. 
We observe that, in all the cases, the binary either undergoes orbital contraction or stalls. 
In the circular case, the presence of radiation pressure slows down the binary inspiral. 
This is consistent with the fact that the accretion torque component $\Delta L_{\rm acc}$ is slightly stronger in the simulations with radiation pressure. 
In the moderate eccentric case $e=0.45$, after a transient phase of $\sim 300$ orbits, the semi-major axis is still approximately equal to $1$. 
In the highly eccentric case $e=0.9$, we find orbital contraction 
regardless of the presence of radiation pressure. 
However, in the radiation pressure case, the binary shrinks between $900-950$ orbital periods and then stabilises, maintaining a nearly constant semi-major axis aroud $\sim 0.997$. Without the radiation pressure, the semi-major axis decreases and still reaches value $ \sim 0.997$, but with some oscillations. 

In the initially circular binary, the eccentricity $e$ experiences a slight increase, reaching values of $\sim 0.003$, independent of radiation pressure presence. In the $e=0.45$ case, the eccentricity increases marginally to $e \sim 0.46-0.47$, while with $e=0.9$ the eccentricity remains essentially constant.
We therefore do not find a unique value at which the binary eccentricity saturates.

The accretion onto the binary is generally lower in the presence of radiation pressure, except during a transient phase of $\sim 400 \, P_{\rm B}$ in the middle eccentric case. 
This result is consistent with the fact that the disc is geometrically thinner and colder if we include radiation pressure in our simulations, see Figure \ref{fig:hr_teff}. 
The radiation pressure contribution causes the accretion rate to decrease from values $\gtrsim \dot M_{\rm edd} $ to average values between $ \dot M \sim 0.1-0.01 \, M_{\odot}/\rm yr$. 


\begin{figure}[h]
    \begin{center}
    \includegraphics[width=\columnwidth]{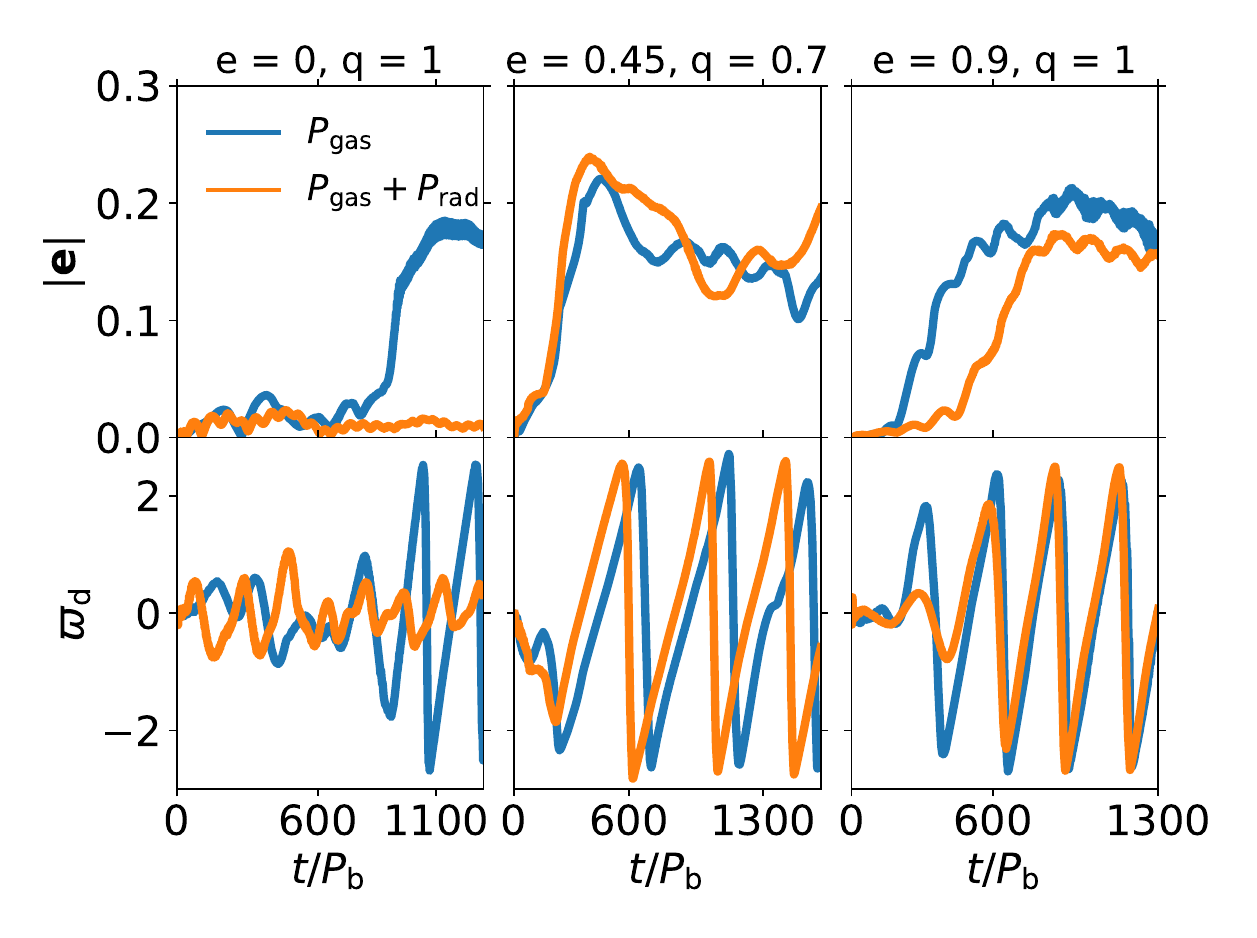} \\   
    \caption{Eccentricity vector evolution over time for the simulations with $e=0$, $q=1$ binaries (left column), $e=0.45$, $q=0.7$ binaries (centre column), and $e=0.9$, $q=1$ binaries (right column).
    The top panels show the magnitude and the bottom panels show the phase in radians. The orange and blue lines refer to the simulation with and without the radiation pressure contribution, respectively. 
    }
    \label{fig:ecc} 

    \end{center}

\end{figure}

\subsection{Interplay between disc shape and binary preferential accretion}

We here investigate the radiation pressure effect on the disc morphology and how this can reflect on the accretion of gas onto the binary components.
We can see from Fig. \ref{fig:disc} that, during the binary evolution, the cavity becomes more eccentric when radiation pressure is excluded from simulations. Additionally, the precession of the cavity also appears to be different between simulations with and without radiation pressure. 

In order to compare the evolution of the cavity eccentricity and its precession 
under different conditions and binary configurations, we compute the Laplace-Runge-Lenz eccentricity vector for each gas element $i$:

\begin{equation}
    \label{eqn:ecc}
     \mathbf{e}_{\rm i} = \frac{|\mathbf{v}_{\rm i}|^2\mathbf{r}_{\rm i}-(\mathbf{v}_{\rm i} \cdot \mathbf{r}_{\rm i})\mathbf{v}_{\rm i}}{GM_{\rm B}}-\hat{\mathbf{r}_{\rm i}}.
\end{equation}
Following this definition, we compute the orbital eccentricity as:

\begin{equation}
    |\mathbf{e}| = \sqrt{e_{\rm x}^2+e_{\rm y}^2},
\end{equation}
where $e_{\rm x}$ and $e_{\rm y}$ are the Cartesian component of the vector.
We determine the longitude of the pericenter of the disc $\varpi_{\rm d} $ as 
\begin{equation}
\label{eqn:phase}
    \varpi_{\rm d}= \arctan \frac{e_{\rm y}}{e_{\rm x}}.
\end{equation}
We compute the mass weighted eccentricity vector between $r=a$ and $r=4a$ of all the gas elements with a bound orbit (e.i. $E_{\rm i}<0$) over 1300 and 1600 orbital periods, for the equal and unequal mass simulations, respectively.

\begin{figure*}
    \begin{center}
    \includegraphics[width=1.9\columnwidth]{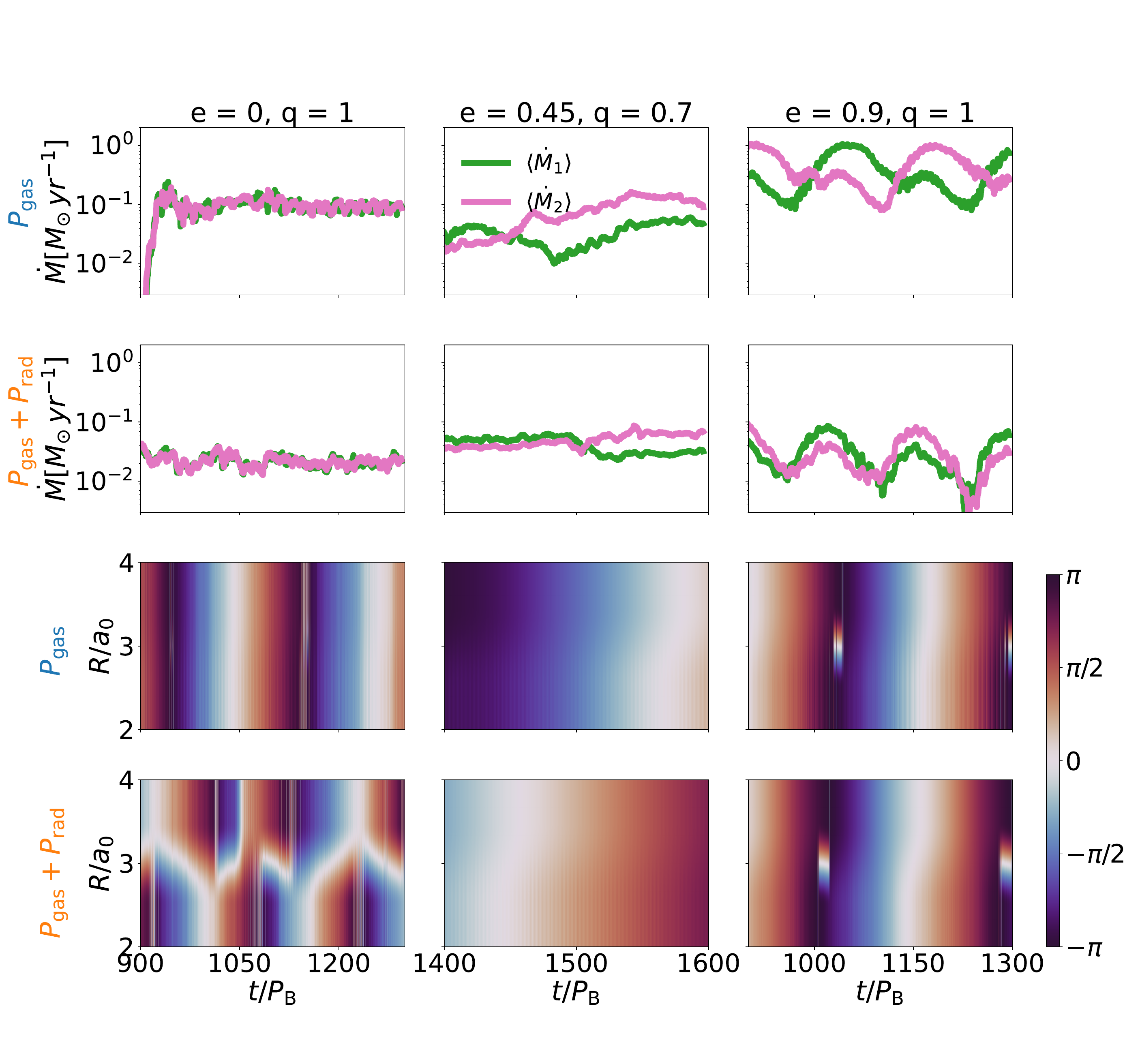}
    \put(5,110){\rotatebox{90}{\color{black} \Large $\varpi_{\rm d} - \varpi_{\rm b}$ }}\\   
    \caption{
    Time evolution of the individual accretion rate (first and second row) and the relative precession phase between the disc and the binary $\varpi = \varpi_{\rm d} - \varpi_{\rm b}$ (third and last row), for the equal mass circular case $e=0$ $q=1$ (left column), $e=0.45$ $q=0.7$ (middle column), and the equal mass highly eccentric one $e=0.9$ $q=1$ (right column). The first two rows show the accretion onto the primary (green line) and the secondary (pink line) black hole, over over the last $400$ and $200$ orbits for the equal and unequal mass cases, respectively. The upper row shows the results of simulations without the radiation pressure, while the second row includes the radiation pressure.
    The third and last row display the evolution of the relative precession phase over the radius. The third row shows results without radiation pressure, and the last row includes it. For $e=0.45$ a transient locking between the disc and the binary eccentricity vectors emerges, leading to episodes of preferential accretion.}
    
    \label{fig:lock_angle} 
    \end{center}

\end{figure*}

Figure \ref{fig:ecc} shows the evolution of the disc eccentricity magnitude (top panels) and the longitude of pericentre (bottom panels) for the six simulations. 
The orange and blue lines refer to the simulations with and without the implementation of radiation pressure, respectively. 
In the circular binary case, when the radiation pressure contributes to the hydrostatic equilibrium of the disc, the cavity remains circular and the precession is not significant, unlike in the case without the radiation pressure. 
In the eccentric binary cases, the radiation pressure effect is less dramatic and the disc cavity becomes eccentric and undergoes precession. The evolution of the the cavity eccentricity is similar in both simulations, with and without the radiation pressure.
This suggests that radiation pressure has an important effect in counteracting cavity eccentricity growth and precession in circular equal mass systems. While the effect of radiation pressure is to significantly reduce the disc temperature and locally isothermal simulations find more eccentric cavities for colder discs, we do not find any disc eccentricity growth in our circular binary simulation in the presence of radiation pressure. The main difference is indeed the inclusion of radiation pressure in our work but we caution that in order to pinpoint the exact origin of this eccentricity suppression we would need to carry out more expensive simulations.

We investigate how the precession of the binary with respect to the precession of the cavity affects the accretion dynamics.
In fact, the angle between the the binary eccentricity vector and the disc cavity eccentricity vector plays a crucial role in determining the amount of preferential accretion experienced by each binary component 
\citep{dunhill2015, DOrazio2021, Siwek2022}. 
When this angle settles to a nearly constant value over time, due to synchronised precession between the binary and the disc, we refer to it as locking angle.
In particular, \cite{Siwek2022} showed that non-zero locking angles which depend on both the binary mass ratio and eccentricity, can enhance or dampen preferential accretion: apsidal misalignment between the binary and the cavity eccentricity vectors, especially at higher eccentricities ($e_{\rm B} > 0.6$) may reduce accretion efficiency, slowing mass ratio growth; conversely, apsidal alignment at lower eccentricities leads to more efficient accretion, promoting mass ratio equalisation. 

We compute the alternation of the accretion rate onto each black hole and the preferential accretion ratio $\lambda = \dot M_{2}/\dot M_{1}$ across all the binary configurations, with and without the implementation of radiation pressure. The results are shown in the first two rows of Fig. \ref{fig:lock_angle}. 

\begin{figure*}
    \begin{center}
    \begin{overpic}[width=\columnwidth]{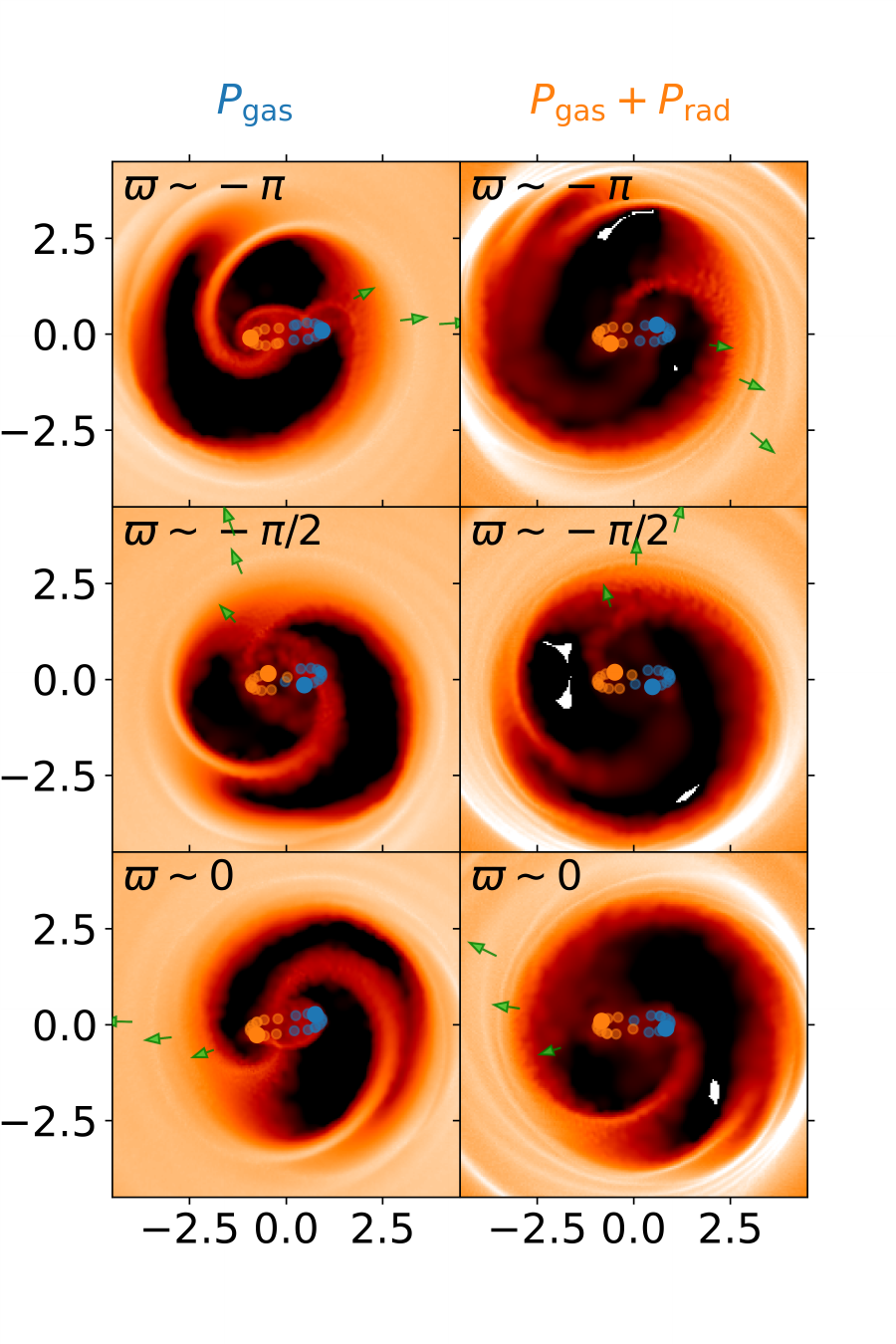}
    \put(25,96){\color{black}\Large e=0.9, q=1}
    \end{overpic}
    \vspace{-0.7cm}
    \begin{overpic}[width=\columnwidth]{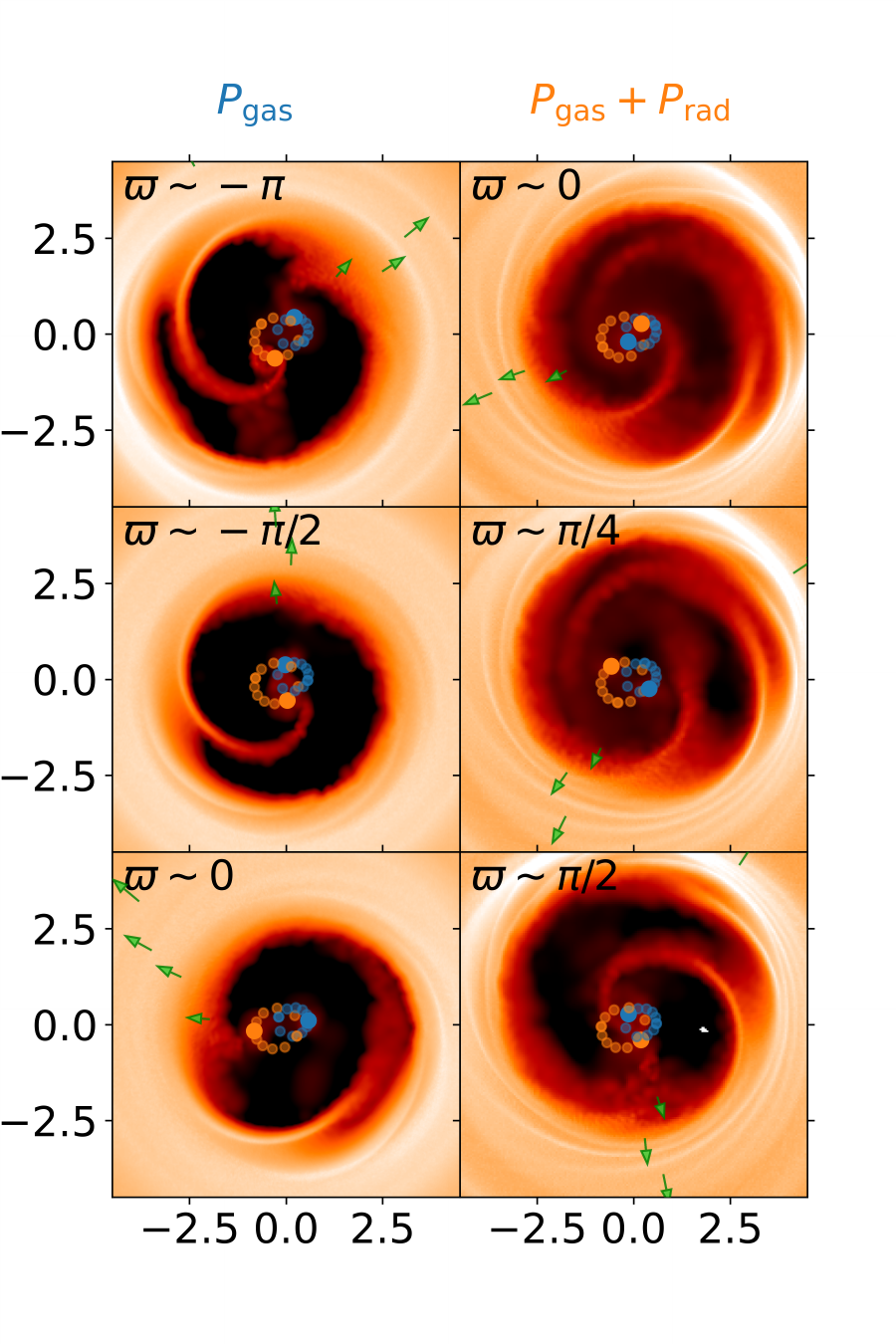}
    \put(23,96){\color{black}\Large e=0.45, q=0.7}
    \end{overpic}
    \caption{Surface density maps in the $x-y$ plane of the circumbinary disc around the equal mass eccentric binary $e=0.9$ and $q=1$ (left panel), and the unequal mass eccentric binary $e=0.45$ and $q=0.7$ (right panel) shown at different relative precession angles. The left and right columns refer to the simulations without and with the radiation pressure, respectively. The solid blue ad orange circles represent the primary and the secondary black hole, respectively, while the semi-transparent circles marks each orbit. The green arrows indicate the disc eccentricity vector phase at increasing radii.}
    \label{fig:diff_angles} 

    \end{center}

\end{figure*}


For the circular equal mass system, the accretion onto the binary components tends to be symmetric. We indeed find $\lambda=1$, both with and without the implementation of radiation pressure, consistent with the expected value from the empirical relation $\lambda_{\rm fit} = q^{-0.9}$ \citep{Duffell2020} 
and in agreement with previous gas-only works \citep[e.g.,][]{farris2014,Munoz2020,Duffell2020,dittaman&ryan2021}. 
For the highly eccentric equal mass binaries $e=0.9$ $q=1$, we obtain $\lambda = 1.2$ and $\lambda = 0.98$ with only gas pressure and with radiation pressure, respectively, broadly consistent with \cite{Siwek2022}.

The most interesting regime to investigate is the unequal mass binary case as prolonged 
preferential accretion is expected to occur in unequal mass systems.
In the intermediate eccentricity unequal mass case, we find $\lambda = 2.28$ and $\lambda = 1.18$ without and with the radiation pressure contribution in the simulations. The stronger preferential accretion we observe in the only gas pressure case, results from favourable locking angles between the binary and cavity eccentricity vectors, which significantly enhances the accretion onto the secondary binary component. We note that \cite{Siwek2022}, for a similar case with $e=0.4$ and $q=0.7$, reported $\lambda \approx 2$.
The inclusion of the radiation pressure again leads to a more balanced accretion rate between the binary components. 

In order to better understand the behaviour of the alternation of the accretion rate onto each black hole and the preferential accretion, we analyse the relative precession phase between the disc and the binary, defined as $\varpi = \varpi_d - \varpi_b$. We computed the binary eccentricity vector $\mathbf{e}_{\rm b}$ and its phase $\varpi_{\rm b}$ following equations (\ref{eqn:ecc}) - (\ref{eqn:phase}). The third and last row of Figure \ref{fig:lock_angle} show the time evolution of this relative precession for all our binary configurations. 
We compute the disc eccentricity vector phase using Eq. (\ref{eqn:phase}).
In the circular equal mass case (left panel), the phase difference between the binary eccentricity vector and the disc eccentricity vector exhibits a periodic behaviour, repeating over $\sim 200$ orbital periods. 
When radiation pressure is included, the disc eccentricity growth is strongly suppressed (see Fig. \ref{fig:ecc}) and the disc experiences a periodic behaviour on shorter timescale. The disc eccentricity remains small and therefore, even with an alternating pattern in the relative angle, this does not affect accretion.

In the highly eccentric equal mass scenario (i.e. $e=0.9$, $q=1$) shown in the rightmost panel of Figure \ref{fig:lock_angle}, the angular difference exhibits a similar behaviour in the gas-only and radiation-pressure simulations. In both cases, $\varpi$ 
precesses over approximately $260$ orbital periods.
In the $e=0.9$ case, the discontinuity between $\pi$ and $-\pi$ combined with the use of circular binning, which is not optimal for highly eccentric configuration, produces some artefacts in the precession phase maps.
In the moderate eccentric unequal mass case (i.e., $e=0.45$ $q=0.7$) shown in the middle panel of Figure \ref{fig:lock_angle}, the relative precession phase evolves more slowly, with a full cycle spanning  $\sim 400$ orbital periods. 
The full cycle duration is determined by analysing the relative precession phase over a longer time window, while Fig. \ref{fig:lock_angle} shows results only in a narrow time window  ($t=1400-1600 \, P_{\rm B}$) consistent with the time window used in other figures.
This results 
in a temporary locking phase, associated to a strong preferential accretion onto the secondary component. In this simulation we indeed find $\lambda \sim 2$.
The inclusion of radiation pressure does not lead to a significant change in the difference between the relative precession phase between the disc and the binary in this case.  

In order to further investigate the dynamics of the streams that provide material to the binary component during the alternation of the accretion rate onto each black hole and the preferential accretion phases, which are essentially determined by the angle between the disc and binary eccentricity vectors, we show the disc surface density maps for the two eccentric binary simulations in Figure \ref{fig:diff_angles}.
From top to bottom, the panels correspond to angles of $\varpi \sim -\pi, -\pi/2, 0$, except for the $e=0.45$ radiation pressure case, where we reported $0, \pi/4, \pi/2$. These configurations respectively represent accretion preferentially onto the primary black hole, symmetric accretion onto both components, and accretion predominantly onto the secondary. 

In the eccentric equal mass case $e=0.9, q=1$, regardless of the presence or not of radiation pressure, when the disc and the binary eccentricity vectors are antiparallel, in agreement with Figure \ref{fig:lock_angle}, the accretion is preferentially onto the black hole located near the cavity edge. When the vectors are perpendicular, the two binary components accrete the same amount of gas. Finally, when the disc and the binary eccentricity vectors are parallel, accretion is favoured onto the black hole further away the cavity edge. 

In the middle eccentric case $e=0.45, q=0.7$, the behaviour is more complex. In the gas pressure only simulation, we find a similar result to the equal mass case: when $\varpi_{\rm d}$ and $\varpi_{\rm b}$ are antiparallel, accretion is preferentially onto the primary black hole, when they are aligned, it favours the secondary. However, the configuration at $\pi/2$ results in a preferential inflow onto the secondary, which is located closer to the cavity edge. 
When the radiation pressure is included, the trend is reversed: accretion occurs preferentially onto the primary when the vectors $\varpi_{\rm d}$ and $\varpi_{\rm b}$ are aligned. As in the only-gas simulation, for $\varpi \sim \pi/2$, the black hole closer to the cavity accretes more gas from the circumbinary disc.



\section{Conclusions}
\label{sec:conclusions}
In this work, we studied the impact of radiation pressure in the evolution of accreting MBHBs at milli-parsec-scales embedded in thin circumbinary gaseous discs using 3D numerical simulations with hyper-Lagrangian refinement. We describe the discs using an adiabatic equation of state, allowing the gas to change its temperature through shocks, PdV work, and black body radiation. We explored binary eccentricities $e=0,0.45,0.9$ and mass ratios $q=1, 0.7$. 
We investigated the role that radiation pressure plays in the evolution of disc aspect ratio $H/R$, surface density $\Sigma$ and effective temperature $T_{\rm eff}$. We additionally measured the torques exerted by the disc onto the binary and the effect they have on the orbital evolution of the binary by directly measuring the change in eccentricity $e$, semi-major axis $a$ and accretion rate. Additionally, we studied the alternation of the accretion rate onto each black hole and the preferential accretion, focusing on its dependence on the angle 
between the disc and the binary eccentricity vectors. 

We find that radiation pressure suppresses the growth of cavity eccentricity and its precession in circular equal-mass binaries, while it has negligible impact in eccentric systems.
In the circular equal mass case the effect of radiation pressure is to effectively suppress disc eccentricity growth and therefore prohibit the formation of the "lump", i.e. over-density at the cavity inner edge.  
Conversely, we find that if the binary is eccentric, the effect of radiation pressure is to lead to a more pronounced lump (see Fig. \ref{fig:disc}). This is consistent with the disc aspect ratio being significantly lower in the simulations with radiation pressure, which, consistently with other studies \citep{Franchini2024a}, contributes to enhancing the over-density contrast. In these cases the high binary eccentricity can overcome the effect of radiation pressure suppressing the cavity eccentricity growth while in the circular binary case radiation pressure maintains the cavity circular.

Radiation pressure plays a crucial role in regulating both the vertical structure and thermal evolution of the circumbinary disc. 
In all simulations, the disc initially cools over the first $\sim 400$ orbits, reducing its aspect ratio from $H/R = 0.1$ to $H/R \sim 0.01$, as it attempts to reach a self-regulated state with Toomre parameter $Q\sim1$. When radiation pressure is included, the disc aspect ratio remains at value $\sim 0.01$ except for the innermost regions, where $H/R \sim 0.05$ due to shocks driven by gas streams that are flung back from the binary and impact the cavity edge. Since the disc is initially completely radiation pressure dominated,  the cooling dominates the heating and the disc collapses in the vertical direction, consistent with previous more sophisticated numerical simulations in the literature \citep{sadowski2016, mishra2016}. The presence of radiation pressure leads to effective temperatures up to a factor of 3 lower than in simulations with only gas pressure. Accretion rate onto the binary are lower when radiation pressure is included ($\dot M \sim 0.01-0.1 \, \rm M_{\odot}/yr$), consistent with the colder, thinner disc structure. 

Since we evolve the binary orbit with time, we can track the evolution of the binary orbital elements and assess the dynamical impact of radiation pressure. We find that radiation pressure does not significantly affect the conservation of total angular momentum.
The gravitational torque exerted by the disc is always negative, consistent with the fact that our circumbinary discs self-regulate towards aspect ratio lower than $0.1$, i.e. the value considered to be the threshold for binary outspiral \citep{SantaBarbara2024}. The torque due to accretion of particles onto the binary components is always positive and slightly higher in the radiation pressure simulations cases. However, its magnitude is never sufficiently large to counteract the negative gravitational torque and we therefore find our binaries to always decrease their semi-major axis. We note that in the very high eccentricity case with radiation pressure the binary undergoes an initial brief phase of angular momentum gain. However, this is not sufficient to drive a meaningful outspiral, as we find in Figure \ref{fig:a_e_mdot}.

The binary eccentricity remains mostly constant, with only minor variations ($e\sim 0.003$ in the circular case and $e\sim 0.46$ in the $e=0.45$ and $q=0.7$ case). We therefore do not find convergence towards a unique value, regardless of the presence of radiation pressure in the simulations. 
While it is possible that the systems have not yer reached an equilibrium eccentricity due to the limited duration of the runs, we note that our results are consistent with those of \cite{Siwek2023}, who similarly find no evolution of the binary eccentricity in the circular equal mass case and an eccentricity growth in the $e=0.4,0.5$ and $q=0.7$ cases over a longer timescale.

We investigate the interplay between the disc morphology and the alternation of accretion  onto each black hole.
In the unequal mass binary case we explored, the radiation pressure contribution reduces the preferential accretion onto the secondary, resulting in a more balanced mass growth between the binary components. In particular, in the gas pressure only eccentric $e=0.45$ simulation, we find the ratio between secondary and primary black hole to be $\lambda = \dot M_2/ \dot M_1 \sim 2.3$, in agreement with \cite{Siwek2022}. However, when radiation pressure is included, the accretion is more balanced and $\lambda = 1.18$ and mass ratio growth is effectively suppressed. In order to better understand the behaviour of the alternation of the accretion rate onto each black hole and the preferential accretion, we analyse the relative precession phase between the disc and the binary defined as $\varpi = \varpi_{\rm d} - \varpi_{\rm b}$. In equal mass binary cases, we find that $\varpi$ cycles over $\sim 200$ and $\sim 260$ orbital periods for eccentricities $e=0$ and $e=0.9$, respectively, with shorter cycles in cases with the radiation pressure. For the eccentric unequal mass binary ($e=0.45$ and $q=0.7$), $\varpi$ evolves more slowly, allowing temporary locking phases between the disc and the binary. This phases correspond to episodes of enhanced accretion onto one binary component over time. In the radiation pressure simulations, accretion occurs onto the secondary black hole when $\varpi \sim 0$ (i.e., disc and binary eccentricity vectors are aligned). For intermediate angles such as $\pi/2$, we find that preferential accretion can still occur in the unequal mass case, where the black hole closer to the cavity tends to accrete more.

Finally, we note that, since radiation pressure in hydrodynamics simulations of MBHBs regulates the thermal structure of the circumbinary disc, it may also impact the electromagnetic emission of MBHBs, possibly affecting the periodic signatures in emitted light used to identify these objects in observations. This is the subject of a follow up work \cite(Cocchiararo-stub-EM).

\section*{Acknowledgements}

We thank Daniel Price for providing the {\sc phantom} code for numerical simulations and acknowledge the use of {\sc splash} \citep{Price2007} for the rendering of the figures.
We thank Phil Hopkins for providing the {\sc gizmo} code for numerical simulations. 
AF and AS acknowledge financial support provided under the European Union’s H2020 ERC Consolidator Grant ``Binary Massive Black Hole Astrophysics" (B Massive, Grant Agreement: 818691). AS also acknowledges the financial support provided under the European Union’s H2020 ERC Advanced Grant ``PINGU'' (Grant Agreement: 101142079). AL acknowledges support by the PRIN MUR "2022935STW".


%
%
\bibliographystyle{aa} 
\bibliography{bibliography}
\end{document}